\documentclass[11pt,floatfix,tightenlines,superscriptaddress]{revtex4-2}
\usepackage{graphicx,graphics}
\usepackage{amsmath,epsf,subfigure}
\usepackage{color}
\usepackage{amsfonts,amsmath,graphics}
\usepackage{bm,subfigure}
\usepackage{graphicx,epsf}
\usepackage{amssymb}
\usepackage{afterpage}
\usepackage{float}
\usepackage{chemmacros}
\usepackage[latin1]{inputenc}
\usepackage[T1]{fontenc}
\usepackage{MnSymbol}
\usepackage{ulem}
\usepackage{verbatim}

\newcommand{\beq}{\begin{equation}}
\newcommand{\eeq}{\end{equation}}
\newcommand{\bea}{\begin{eqnarray}}
\newcommand{\eea}{\end{eqnarray}}

\renewcommand{\a}{\alpha}

\renewcommand{\d}{\delta}
\newcommand{\e}{\epsilon}

\newcommand{\s}{\sigma}

\renewcommand{\p}{\partial}

\renewcommand{\L}{\Lambda}
\newcommand{\G}{\Gamma}

\renewcommand{\S}{\mathcal{S}}

\begin{document}

\title{Learning to outgrow the competition: reaction-diffusion systems \\ that adapt to time-dependent environments}
\author{Olivier Rivoire}
\affiliation{Gulliver, CNRS, ESPCI, Universit\'e PSL, 75005 Paris, France}
\author{Guy Bunin}
\affiliation{Department of Physics, Technion-Israel Institute of Technology, Haifa 32000, Israel}

\begin{abstract} 
A fundamental challenge in both biology and engineering is understanding how adaptation can emerge from simple physical or chemical building blocks. Biological adaptation appears to rely on two key capabilities: information processing, to enable learning of complex tasks, and reproduction, to give such learning its value through reproductive success. Creating and coordinating these capabilities poses both a formidable engineering challenge and a fundamental evolutionary puzzle.
Here, we propose a model in which learning and reproduction emerge jointly from simple, nonequilibrium chemical systems, without prior design or evolution. 
The model's essential features are that it supports many steady states and spatial heterogeneities, upon which adaptation arises solely from exposure to time-varying influxes of reactants, without any predesigned program or memory. The adaptive response is specific to the temporal sequence of environmental changes, yet general enough to extend to related sequences.
Learning manifests in several forms: it is reinforced through repeated exposure to the same environmental sequence, representing a form of self-learning, and enhanced through spatial interactions, corresponding to a form of collective learning. Moreover, an adapted state can accelerate adaptation in nearby regions, providing a mechanism akin to teacher-guided learning.
By coupling environmental variations to stochastic reaction-diffusion dynamics, our model establishes a  minimal physical framework in which learning of complex environments and selective amplification of the states that encode them, arise directly from nonequilibrium chemical processes.
\end{abstract}

\maketitle

A hallmark of living systems is their ability to adapt to complex, changing environments. How does such behavior arise in biology, and how can it be engineered in artificial systems? While even the simplest organisms exhibit adaptation, they already represent highly intricate systems, with adaptive mechanisms that are far from fully understood~\cite{gershman2021reconsidering}.
This drives efforts to find general principles of adaptation beyond contemporary biology, both to shed light on prebiotic origins of adaptive processes, and to guide the engineering of adaptive systems~\cite{goldenfeld2011life,furusawa2012adaptation,perunov2016statistical,landmann2021simple,trifonova2025trainable}. Adopting a minimal, bottom-up perspective, the aim is to obtain rich, biological-like adaptation using only simple chemical or physical components, such as short molecules, small RNAs, or colloids~\cite{lehn2015perspectives,van2015programmable,zeravcic2017colloquium,adamski2020self,pavlinova2023abiogenesis}. It is this significant challenge  that we address in this work.

A fundamental difficulty is that biological adaptation seems to require two key abilities: information processing to enable adaptive responses to environmental stimuli, and reproduction to express this adaptation through growth. Indeed, biological adaptation is ultimately defined by whether it leads to descendants that outgrow those of others.
Considerable efforts in engineering and theory have gone into creating simple physical or chemical systems capable of learning~\cite{stern2023learning,nagipogu2023survey,paulsen2025mechanical, cherry2025supervised}, which involves information processing, and to designing simple self-replicators~\cite{sipper1998fifty,schulman2012robust,zhuo2019litters,adamski2020self}. The two dominant frameworks for these goals are machine learning for learning and autocatalysis for replication.
Yet each of these goals still presents challenges, and current frameworks offer no straightforward path to integrating the two capabilities.
From a biological standpoint, the dominant framework to understand adaptation is evolution by natural selection, but it presents a chicken-and-egg conundrum: since information processing and reproduction are prerequisites for evolution, they cannot have arisen through evolution itself.

Here, we propose a paradigm in which advanced learning and reproduction can emerge together from simple chemical systems, enabling complex adaptation. Unlike current frameworks, our approach does not rely on machine learning, autocatalysis, or natural selection, thereby overcoming their inherent limitations. Instead, our framework leverages systems with multiple non-equilibrium steady states and spatial heterogeneity, so that local regions can adopt distinct steady states and interact through diffusion~\cite{bunin2025evolutionary}.

This framework does not involve explicit engineering of reproduction or learning, yet gives rise to states that can adapt in complex environments. Furthermore, it exhibits distinct modalities of adaptation, reminiscent of different forms of biological adaptation: self-adaptation through direct interaction with the environment, even in the absence of reproduction (analogous to physiological adaptation), and additional adaptation when reproduction through spatial expansion is enabled (analogous to evolutionary adaptation). This second modality can be interpreted as ``collective learning'' among a population of ``students'', and is further enhanced by the presence of a ``teacher''. Our paradigm thus offers both a fresh engineering perspective on chemical learning and fresh conceptual insights into biological adaptation.

\section*{A general definition of adaptation}

\begin{figure}[t]
\centering
\includegraphics[width=.6\linewidth]{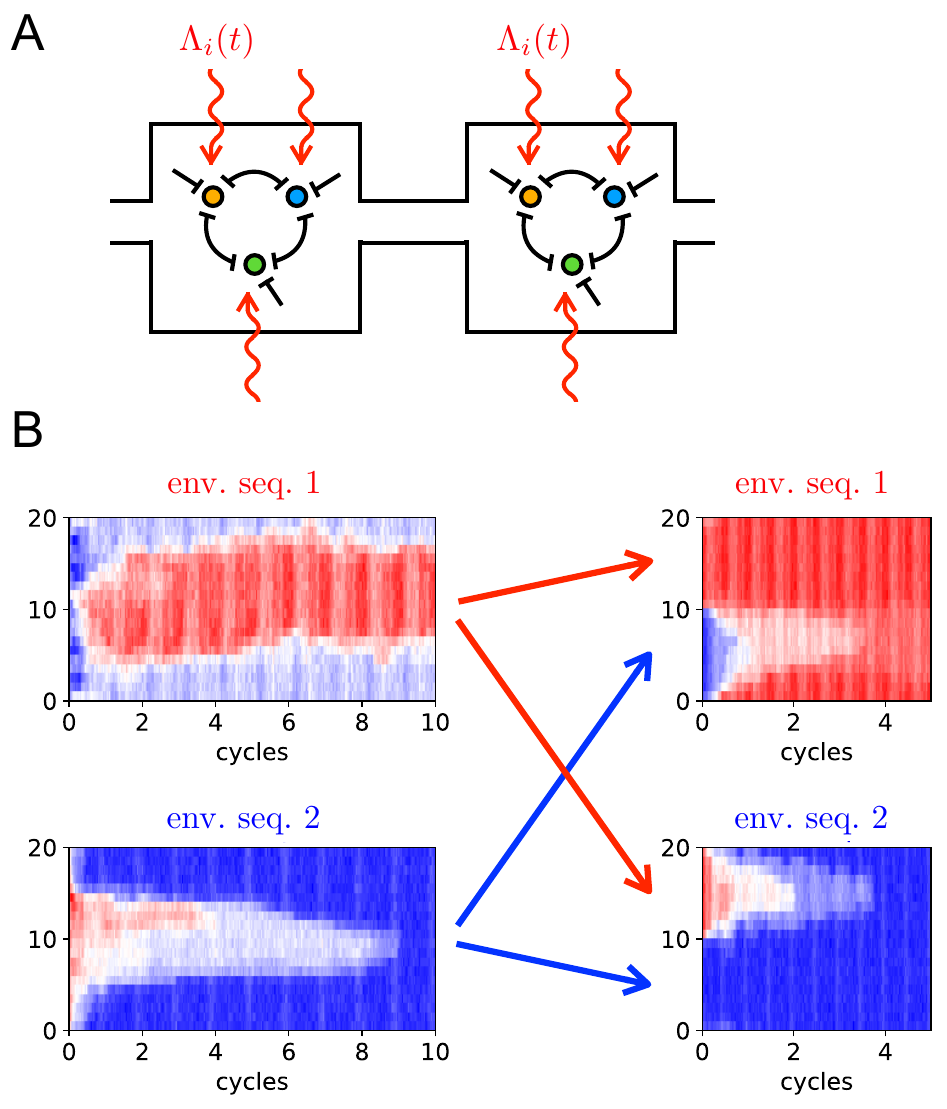}
\caption{{\bf A.} Our definition of adaptation applies to reaction-diffusion systems subjected to time-varying drives. Here, this is represented schematically, with reactions coupled by reciprocal inhibition in a series of reactors connected through diffusion.  Time-varying influxes $\L_i(t)$ of reactants define the environmental drive while uniform dilution (not represented) ensures that reaction volumes are kept constant. {\bf B.}~Protocol for demonstrating adaptation. Left: First (training) stage, where the same system of reactions is subjected to two different periodic drives (here for over 10 cycles). The x-axis represents time, and the y-axis a one-dimensional space of coupled reactors. The densities $n(x,t)$ are colored based on their proximity to two of the final states, with red indicating closeness to a final state with the first environmental drive (top), and blue to one with the second (bottom). Right: Second (testing) stage, where these final states are competed against each other under repeated exposure to each environmental drive. Initially, half of the space is seeded with one final state and the other half with the other. After a few cycles (here 5), all reactors contain states close to the final state corresponding to their respective environmental sequence. When comparable results are obtained for most repetitions of the first and second stages, we say that the system adapts to the environmental drives. }\label{fig:compet}
\label{fig:illustr}
\end{figure}

We first present our definition of adaptation to time-varying environments in a general context before demonstrating how it arises in a particular system. This context is that of stochastic reaction-diffusion systems (Fig.~\ref{fig:illustr}A) described by an equation of the form
\beq\label{eq:rde}
\partial_t n(x,t)
= F\!\left(n(x,t),\epsilon(t)\right)
+ D \nabla^2 n(x,t)
+ {\rm noise}(x,t) .
\eeq
Here $n(x,t)$ is an $N$-component field whose component $n_i(x,t)$ denotes the density of chemical species $i$ at location $x$ and time $t$; $F(n(x,t),\epsilon(t))$ specifies the reaction kinetics, with a time-dependent parameter $\epsilon(t)$ representing the environment; $D\nabla^2 n(x,t)$ is the usual diffusion term; and ${\rm noise}(x,t)$ accounts for stochasticity due to the finite number of molecules, as in chemical Langevin equations (see Methods for explicit expressions). We consider time-varying environments consisting of periodic drives $\epsilon(t)$ satisfying $\epsilon(t+\tau)=\epsilon(t)$ with a period $\tau$. 

To define adaptation, we propose a protocol consisting of two stages (see Fig.~\ref{fig:illustr}B): a first stage in which the system is allowed to adapt, and a second stage in which this adaptation is tested via competition. In the first (training) stage, the system experiences a particular periodic drive $\epsilon_1(t)$ for $k_{\rm train}$ periods, starting from an initial condition $n_0(x)$. We then select at random a location $x^*$ and extract the final configuration at this location, $n_1 = n(x^*,k_{\rm train}\tau)$. For comparison, the same system starting from the same initial condition is subjected to a different periodic drive $\epsilon_2(t)$ with the same period $\tau$ for the same total duration, and we retain the final state $n_2 = n(x^*,k_{\rm train}\tau)$ at the same location $x^*$. In the second (testing) stage, we divide the spatial domain $S$ into two equivalent subdomains $S_1$ and $S_2$, and initialize each with one of the two previously obtained states, i.e.\ $n(x,0)=n_1$ if $x\in S_1$ and $n(x,0)=n_2$ if $x\in S_2$. We then run the dynamics for $k_{\rm test}$ periods under one of the two periodic drives, say $\epsilon_1(t)$. Finally, we say that the system has adapted to  $\epsilon_1(t)$ if the final states $n(x,k_{\rm test}\tau)$ for $x\in S_2$ are more similar to $n_1$ than to $n_2$, on average over many realizations of the training and testing stages.

\section*{Model: Single reactor}

\begin{figure}[t]
\centering
\includegraphics[width=.7\linewidth]{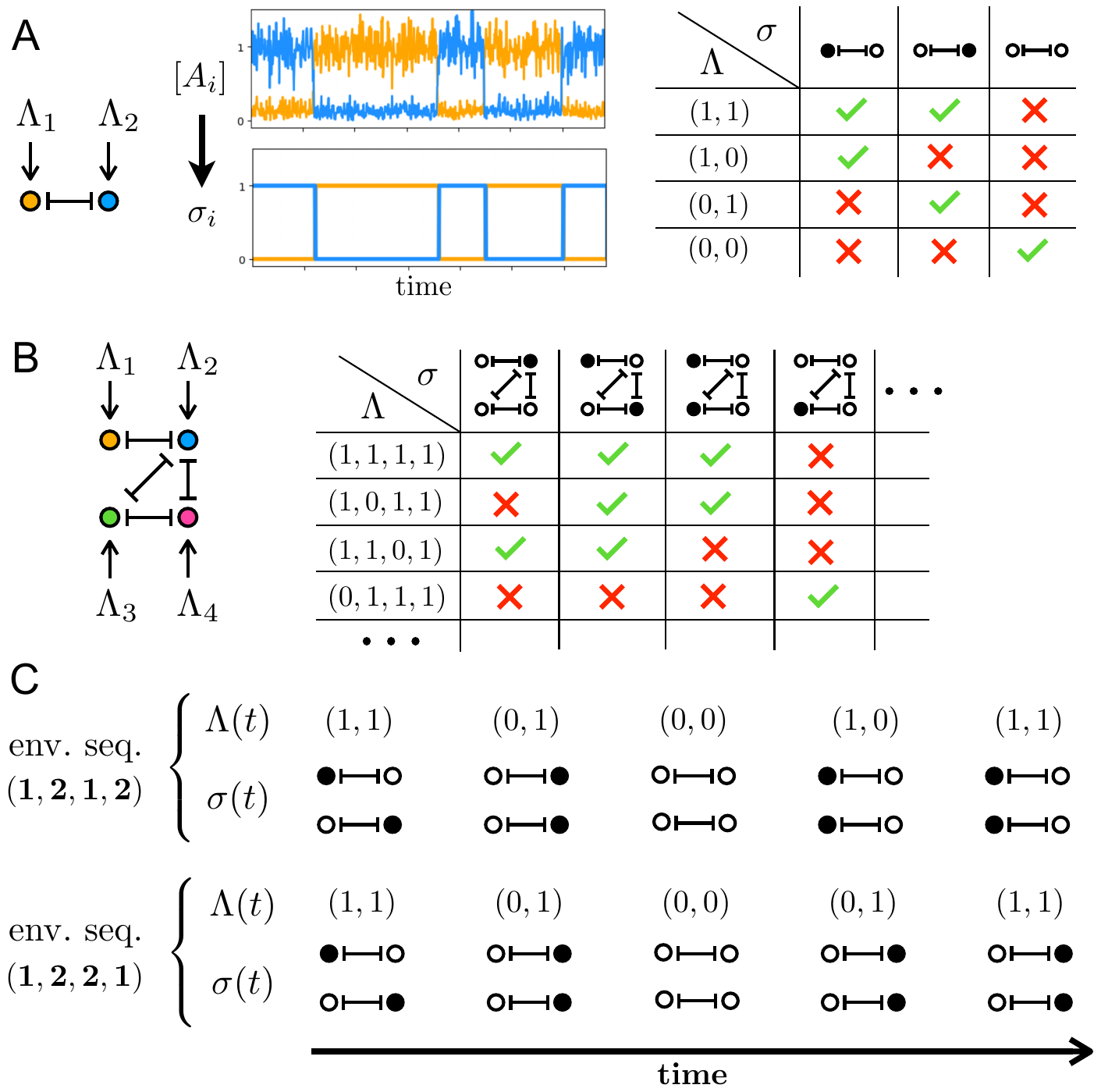}
\caption{Systems, states and environmental sequences. A system is
specified by $N$ nodes, with mutual inhibitions between some pairs
of nodes, represented by $\vdash\!\dashv$. The environment $\Lambda$
and the state of the system $\sigma$ are $N$-dimensional vectors
with elements in $\{0,1\}$. Dynamically, a node becomes active, $\sigma_i=1$,
when its influx is turned on, $\Lambda_i=1$, and no other node that inhibits it is active; a node becomes inactive as soon as its
influx turns off, to $\Lambda_i=0$. 
{\bf A.} A system with $N=2$ nodes, with an inhibitory interaction. When all influxes are on, i.e., $\L=(1,1)$, the system can be in two states, $\s=(1,0)$ or $\s=(0,1)$. Top graph: continuous dynamics, showing stochastic switches between the two states; bottom: the binary model used here. As indicated in the table, turning off one of the fluxes eliminates the possibility of one of the states. Here, open circles represent inactive nodes and full circles active nodes.
{\bf B.} The same principles apply to larger networks combining multiple nodes and interactions between them, where more states are present and where the relation between the influxes and the states is more intricate, as illustrated here with a network of $N=4$ nodes.
{\bf C.} An environmental sequence is defined by a series of $2m$ nodes $(i_1,\dots,i_{2m})$ where each node appears an even number of times. It represents a series of events where fluxes $\L_i$ are successively turned either on or off, starting and ending with all fluxes on. 
For example, in the environmental sequence $(1,2,1,2)$, $\L_1$ is turned off, followed by $\L_2$, before $\L_1$ is turned on, followed by $\L_2$. Here, the dynamics of the system with $N=2$ is represented when experiencing two particular environmental sequences, $(1,2,1,2)$ and $(1,2,2,1)$, starting each time from the two possible initial conditions, $\s=(1,0)$ and $\s=(0,1)$. In this example, the final state depends on the environmental sequence while the initial state is forgotten.
}\label{fig:scheme}
\end{figure}

We will demonstrate that adaptation as defined in the previous section can arise in a specific system of reactions when subjected to a class of environmental sequences. First, we define this system and these environmental sequences.

\subsection*{Reaction dynamics}

We start by describing the reactions in the context of a single, well-mixed reactor. The key ingredients are $N$ molecules $A_i$ ($i=1,\dots,N)$ whose concentrations $[A_i]$ define the components $n_i$ of the $N$-dimensional field $n$ in \eqref{eq:rde}. Each $A_i$ is produced from a corresponding resource $R_i$ when this resource is injected in the reactor.
The $N$ possible reactions of formation of $A_i$ from $R_i$ are coupled by sparse reciprocal inhibitions: When the product $A_i$ of reaction $i$ is present at high concentration, it prevents some other reactions $j$ from occurring, and therefore the associated $A_j$ from being produced. The simplest case with $N=2$ reactions is illustrated in Fig.~\ref{fig:scheme}A. One possible mechanism is for the $A_i$ to inhibit the catalyst required for forming $A_j$ from $R_j$ (see Methods). The inhibitions are sparse, meaning that each $A_i$ inhibits only a few reactions $j$, and they are reciprocal meaning that whenever $A_i$ inhibits reaction $j$, $A_j$ inhibits reaction $i$ (denoted as $\vdash\!\dashv$).

These reactions take place in an open reactor in which each reactant $R_i$ is supplied with an influx $\L_i$ and where constant dilution ensures a fixed volume. At any given time, each reactant $R_i$ is either supplied ($\L_i=1$) or not supplied ($\L_i=0$). In the simple case with $N=2$ reactions (Fig.~\ref{fig:scheme}A), when both $R_1$ and $R_2$ are supplied, corresponding to the influx vector $\L=(1,1)$, two stable steady states are possible: one where $A_1$ is high in concentration and $A_2$ is low, and another where this is reversed. These concentrations are generally continuous but here we adopt a binary description valid at low noise (Fig.~\ref{fig:scheme}A and Methods) where one state is represented by $\s=(1,0)$ and the other by $\s=(0,1)$. If instead only $R_1$ is supplied, corresponding to $\L=(1,0)$, only the first state $\s=(1,0)$ is possible (see table in Fig.~\ref{fig:scheme}A). 

This generalizes to larger networks of reactions, represented by a graph whose nodes are the reactions $i$ and whose edges indicate reciprocal inhibitions (Fig.~\ref{fig:scheme}B). When $N$ is large, a very large number of steady states is possible. In the case of a single well-mixed reactor of large volume where stochastic fluctuations are negligible, these stable steady states are characterized by a vector of binary variables $\s\in\{0,1\}^{N}$ that satisfy 
\beq\label{eq:Constraints}
(1-\L_i)\s_i=0,\quad\forall i\quad{\rm and}\quad \s_{i}\s_{j}=0,\quad\forall i\vdash\!\dashv j.
\eeq
The first constraint imposes that $\s_i=0$ whenever $\L_i=0$, and the second that $\s_{i}$ and $\s_{j}$ cannot both be 1 when reactions $i$ and $j$ reciprocally inhibit each other. 

Noise due to the limited number of particles in reactors of finite volume can cause switches between the possible steady states. While the dynamics of these switches is fundamentally out-of-equilibrium, it is formally equivalent to the equilibrium dynamics of the thermal system with energy
\beq\label{eq:Energy}
E[\s]=-\sum_{i=1}^N \s_{i}
\eeq
subject to the conditions of \eqref{eq:Constraints},
and with temperature $T=1/\sqrt{\Omega}$, where $\Omega$ is the finite volume of the reactor. 
This equivalence is permitted by the reciprocity of the inhibitions and is valid under a low-noise condition (see Methods). 

We introduce time-dependent drives by varying the influxes $\L_i$. The time $\tau$ between two changes is chosen to be long enough to allow the system to relax to the nearest stable steady state following each environmental shift, yet short compared to the timescale required for full relaxation of the entire system. The stochastic dynamics is simulated using a kinetic Metropolis Monte Carlo algorithm.

\subsection*{Time-dependent environmental sequences}

We consider environmental sequences where $\L(t)$ changes by turning on or off one resource $R_i$ at a time (Fig.~\ref{fig:scheme}C). Initially, all resources are supplied, with $\L_i=1$ for all $i$. The first step consists in turning off the supply of $R_{i_1}$ by setting $\L_{i_1}$ to $0$. Recursively, the $k$th step consists in setting $\L_{i_k}$ to 0 if it was 1, and to 1 if it was 0. We perform $2m$ such steps, where each $i_k$ appears an even number of times (possibly zero) so that after the last step, all the $\L_i$ have returned to 1. 
Thus, a given environmental sequence is represented by a series $(i_1,\dots,i_{2m})$ of reactions, where each reaction appears an even number of times.
After each of the $2m$ steps, the system relaxes to the nearest stable state. These series of environmental changes generate transitions between states, as illustrated in Fig.~\ref{fig:scheme}C. 

In the following figures, we consider $N=50$ reactions with a pattern of reciprocal inhibition given by a random regular graph of connectivity $3$. Unless otherwise mentioned, we show results averaged over 10 random graphs and, for each graph, over 1000 independent trajectories, all starting with an empty state where $\s_{i}=0$ for all $i$. Below, we generalize the model by considering multiple coupled reactors.

\section*{Encoding of temporal sequences}

\begin{figure}[t]
\centering
\includegraphics[width=.7\linewidth]{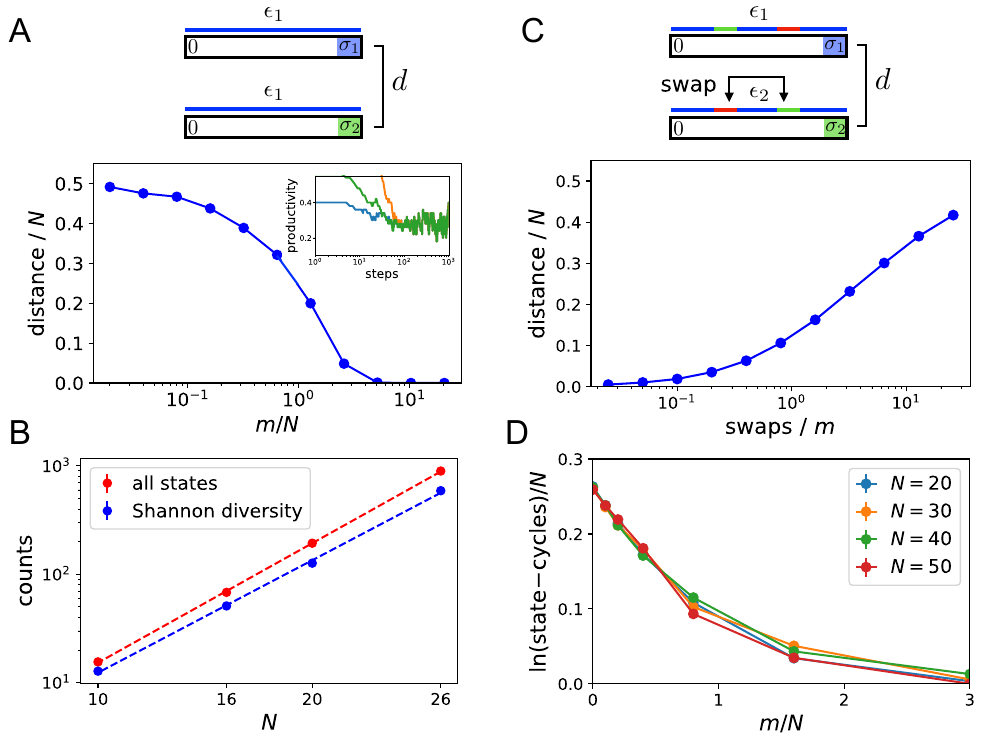}
\caption{Encoding of information from environmental sequences in final states. {\bf A.}~Mean distance between states reached after experiencing the same environmental sequence for varying values of $m$. Two trajectories are run under the same environmental sequence and the distance (normalized Hamming distance) between the two states at the end is reported. For $m$ sufficiently large ($m/N\gg 1$) states obtained after the same environmental sequence are identical. Inset: mean number of active reactions (productivity) as a function of the number of steps with $m=500$, starting from 3 different initial conditions, showing convergence. {\bf B.}~Number of different states reached after a long environmental sequence ($m=1000$) when sampling uniformly environmental sequences, measured as the Shannon diversity, with an exponential fit, shown as a function of the number $N$ of reactions. {\bf C.} Distance between states obtained from two environmental sequences differing by a given number of swaps, with $m=1000$. {\bf D.} Mean number of different state-cycles for a given environmental sequence (from exhaustive search).}\label{fig:sequences}
\end{figure}

First, we look at states obtained after applying an environmental sequence, and study how these states encode information on the sequence. To focus on the encoding effects, we set $T=0$ (no noise), so that when some $\L_i$ are turned on or off by setting them to 1 or 0, the state evolves to one of the nearest local minima of the energy function given by \eqref{eq:Energy}-\eqref{eq:Constraints}. As only one $\L_i$ is turned on or off at a time, the dynamics are deterministic provided the graph of inhibitory interactions does not contain triangles, as we consider in this section.

We fix an environmental sequence with a given $m$ and, starting from different initial conditions, record the states that we obtain after completing the environmental sequence. When $m/N\gg 1$, we obtain a single state independent of initial conditions (Fig.~\ref{fig:sequences}A). In this case, a large number of different final states is reached when sampling uniformly different environmental sequence. This number is exponential in $N$ and not far from the total number of states when all $\Lambda_i=1$ (Fig.~\ref{fig:sequences}B). In addition, the encoding is continuous, with similar environmental sequences mapping, on average, to similar states: the distance between final states varies gradually with the number of swaps by which two environmental sequences differ, where one swap from environmental sequence $(i_1,\dots,i_{2m})$ consists in randomly choosing $1\leq k<k'\leq 2m$ and swapping $i_k$ and $i_{k'}$ (Fig.~\ref{fig:sequences}C). When $m/N\lesssim 1$, repeating successively the same environmental sequence causes the system to converge to periodic dynamics, repeating the same sequence of states with each environmental period. The number of such state-cycles associated with a given environmental sequence scales as $e^{N\phi(m/N)}$ where $\phi(m/N)$ decreases to 0 as $m/N$ increases (Fig.~\ref{fig:sequences}D).

\section*{Competitive advantage in coupled reactors}

To analyze whether the information from past environments encoded in a state confers an adaptive advantage, we must quantify whether a state can reproduce more effectively than another. We enable reproduction by introducing spatial extension in the form of multiple reactors coupled by diffusion. This setup allows reproduction in the sense that a state initially confined to a subset of reactors can spread to other reactors. Reproduction here does not require strict replication; rather, it necessitates heredity whereby newly formed states must be correlated with their parent. Since a state is defined by which reactions are active or inhibited, this correlation is measured by the fraction of reactions that are similarly active or inhibited in both parent and offspring states. Using this capacity to reproduce, two states can be competed against each other by seeding them in neighboring reactors. If, after some time, the reactors each contain states more similar to the first than the second, the first state is considered to have reproduced at the expense of the second, and thus to be better adapted to the environment in which the competition occurred. 

\subsection*{Model: reaction-diffusion dynamics}

Specifically, we consider coupling through diffusion $M>1$ reactors arranged along a closed one-dimensional chain. The overall state of the system is represented by an $M \times N$ matrix $\s \in \{0,1\}^{M \times N}$, whose dynamics are governed by an energy that generalizes \eqref{eq:Energy},
\beq\label{eq:multiple}
E[\s]=-\sum_{x=0}^{M-1}\sum_{i=1}^N \left[\s_{x,i}+J\s_{x,i}\s_{(x+1),i}\right]
\eeq
where $x+1$ is taken modulo $M$. The activation and inhibition constraints apply as before within each reactor $x$,
\beq\label{eq:multconstraints}
(1-\L_i)\s_{x,i}=0,\quad\forall x, i\quad{\rm and}\quad \s_{x,i}\s_{x,j}=0,\quad\forall x, i\vdash\!\dashv j.
\eeq
The added spatial coupling term in \eqref{eq:multiple} corresponds to diffusive dynamics in space; indeed, for continuous variables $\sigma_i$ at temperature $T=0$ where the dynamics are gradient descent in energy, this added energy term contributes to the dynamics a term of the form $d\sigma_{x,i}/dt=-\partial E/\partial\s_{x,i}=J[\sigma_{(x+1),i}+\sigma_{(x-1),i}-2\sigma_{x,i}]$, which describes diffusion in discrete space  (see Methods for discussion of the model with binary variables).  

We start by considering two reactors coupled without periodic boundary conditions,
\beq\label{eq:J}
E[\s]=-\sum_{i=1}^N\left[ \s_{1,i}+\s_{2,i} + J\s_{1,i}\s_{2,i}\right]
\eeq
For large values of $m$, the environment uniquely determines the final states (Fig.~\ref{fig:sequences}A), making the outcome of competitions independent of previous experience. To avoid this trivial case of adaptation, we set $m=40$, which allows for multiple state-cycles ($m/N=0.8$ given $N=50$, see Fig.~\ref{fig:sequences}D). We also set the duration of each environmental step to $10$ units of time and the temperature to $T=0.15$ to obtain approximately one state-cycle change every 10 environmental cycles that repeat a particular environmental sequence. Finally, we set $J=0.5$.

\subsection*{Competition and quantifying adaptation}

\begin{figure}[t]
\centering
\includegraphics[width=.7\linewidth]{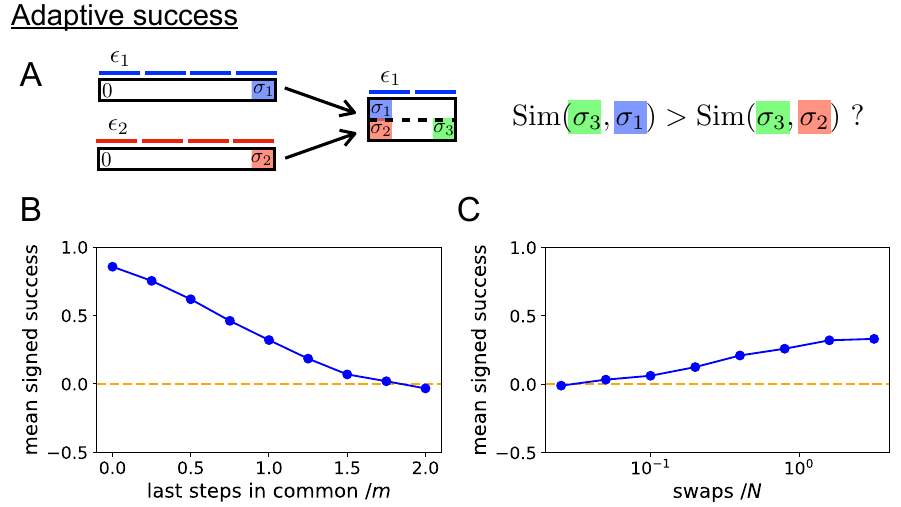}
\caption{Competition and adaptation. {\bf A.} Two states obtained in independent reactors under cycles of two distinct environmental sequences $\e_1$ and $\e_2$ are competed in two coupled reactors under cycles of environmental sequence $\e_1$. Signed success is the fraction of times the state $\s_3$ resulting from the competition is more similar to $\s_1$, the state evolved under the same environment, minus the fraction of cases where it is more similar to $\s_2$.  {\bf B.} The two environmental sequences $\e_1$ and $\e_2$ share a number of last steps in common. Given that there are $2m$ steps, they are identical when steps$/m=2$. {\bf C.} The environmental sequences of the two environments differ by $n$ swaps, showing ability to discriminate changes in the order of steps within an environmental sequence.}\label{fig:adaptA}
\end{figure}

To assess the level of adaptation, we rely on competitions between states obtained after cycles of different environmental sequences, as discussed before (Fig.~\ref{fig:illustr}). Given two environmental sequences $\e_1$ and $\e_2$, we run $c$ cycles of environmental sequence $\e_1$ in one isolated reactor, resulting in state $\s_1$, and $c$ cycles of environmental sequence $\e_2$ in another isolated reactor, resulting in state $\s_2$, each an $N$-dimensional vector in $\{0,1\}^N$. By concatenating these vectors into a $2\times N$ matrix $(\s_1, \s_2)$, we construct the initial condition for a system with two coupled reactors, on which we run $c'$ cycles of another environmental sequence $\e'_1$, either identical  to $\e_1$, or related to it (as in Fig.~\ref{fig:general}B). This yields the pair of states $(\s_4, \s_3)$ (see Fig.~\ref{fig:adaptA}A). By default, we set $c=10^3$ and $c'=10$.

Finally, we compare the similarity of $\sigma_3$ to $\sigma_1$ and to $\sigma_2$, in order to assess the impact of an adapted state in a neighboring reactor. We define the signed difference as $1$ if $\s_3$ is closer to $\s_1$ than to $\s_2$ in terms of the Hamming distance; $-1$ if it is closer to $\s_2$; and $0$ if it is equally similar to both. This quantity is then averaged over reaction networks, environmental sequences and stochastic trajectories to define a mean signed success. 

\begin{figure}[t]
\centering
\includegraphics[width=.45\linewidth]{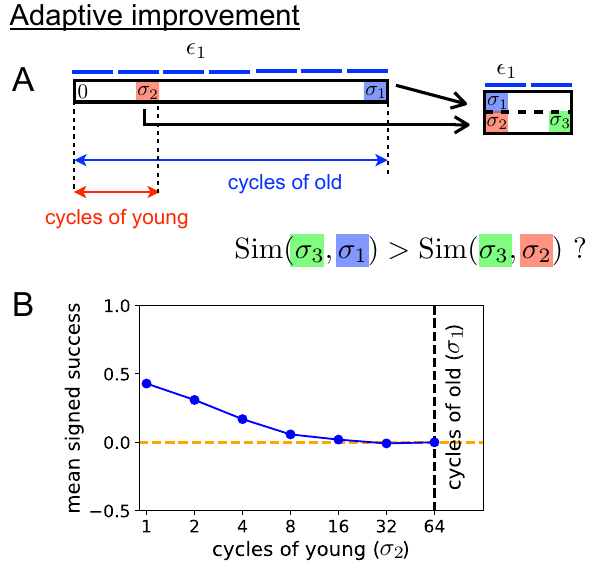}
\caption{Adaptive improvement. {\bf A.} Two competing states are obtained from the same trajectory at different times, a young state $\s_1$ and an old state $\s_2$. {\bf B.} Here the old state is taken after $64$ cycles, the young one from a varying number of cycles (horizontal axis). }\label{fig:adaptB}
\end{figure}

\section*{Adaptation without reproduction}

Assessing whether a state is adapted requires competition through reproduction, which in our model involves multiple reactors that are coupled by diffusion. Yet obtaining an adapted state does not necessarily require reproduction. As we now demonstrate, it can arise in a single well-mixed reactor, before its adaptation becomes manifest in coupled reactors.

\subsection*{Adaptive success}

We first consider competitions under cycles of an environmental sequence identical to that in which one of the competitors evolved, namely $\e_1' = \e_1$. When $\e_2$ is unrelated to $\e_1$, we observe that the signed success, averaged over different reaction networks, environmental sequences and trajectories, is close to $1$ (Fig.~\ref{fig:adaptA}B, leftmost point). This indicates that the final state in reactor $2$ is almost always more similar to the initial state in reactor $1$ than to that in reactor $2$. This observation also holds when the two environments $\e_1$ and $\e_2$ are more subtly different, such as when they share several last steps (Fig.~\ref{fig:adaptA}B), or differ by a finite number of swaps (Fig.~\ref{fig:adaptA}C). 

\subsection*{Adaptive improvement}

One may wonder if the results shown in Fig.~\ref{fig:adaptA}A,B,C can be attributed to the response to the last cycle only, independent of earlier environmental history. From a biological standpoint, one would like to differentiate between such a scenario, which is a basic form of plasticity, and learning, which is distinguished from other forms of plasticity by the fact that it improves upon repeated exposures. Here we demonstrate that such learning indeed happens by competing two states, $\s_1$ and $\s_2$, taken along the same trajectory but differing in age (Fig.~\ref{fig:adaptB}A). We find that, on average, the older state has an adaptive advantage over the younger one (Fig.~\ref{fig:adaptB}B).

\subsection*{Generalization}

\begin{figure}[t]
\centering
\includegraphics[width=.6\linewidth]{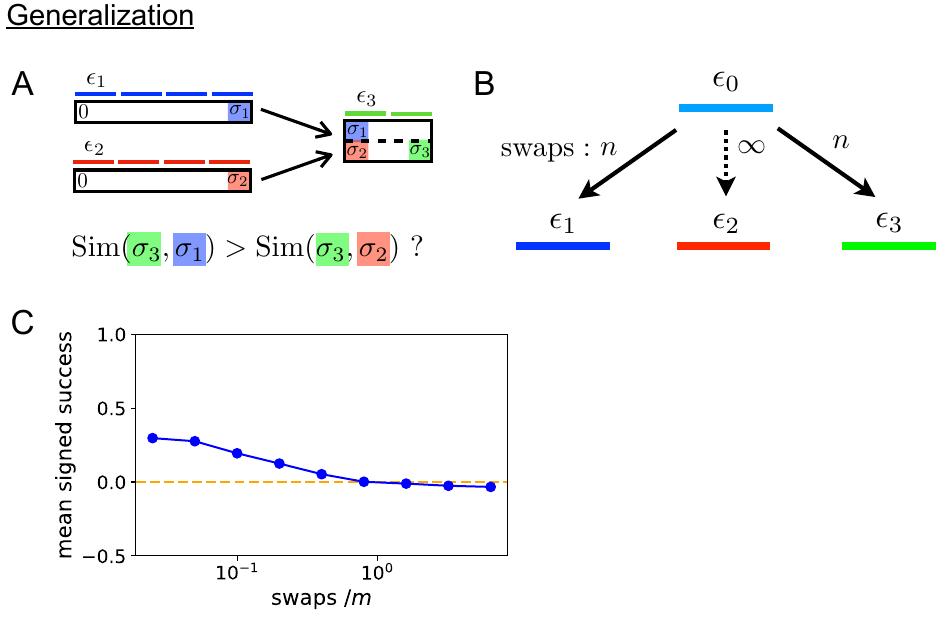}
\caption{Generalization. {\bf A.} Same scheme as Fig.~\ref{fig:adaptA}A but {\bf B.}  the environmental sequences applied during the competition differs from the two used in the training phase. The environmental sequences $\epsilon_1$ and $\epsilon_3$ are independently derived by $n$ swaps from a reference $\epsilon_0$, while the environmental sequence  $\epsilon_2$ is an independent random permutation of $\epsilon_0$. {\bf C.} Generalization success as a function of $n/m$.
}\label{fig:general}
\end{figure}

From another perspective, that of machine learning, learning is a form of generalization. In this context, training involves examples drawn from a probability distribution, and testing other examples from the same distribution. To demonstrate this type of generalization in our model, we repeat the previous competition scheme (Fig.~\ref{fig:general}A), but now define a master environmental sequence $\e_0$ from which $\e_1$ and $\e_3$ are independently generated by performing a finite number of swaps, while $\e_2$ is obtained from $\e_0$ by an infinite number of swaps, representing an arbitrary permutation (Fig.~\ref{fig:general}B). The results show that adaptation does occur in this setting that requires generalization (Fig.~\ref{fig:general}C).

\section*{Enhanced adaptation through interactions}

\subsection*{Coupling to an adapted state}

\begin{figure}[t]
\centering
\includegraphics[width=.45\linewidth]{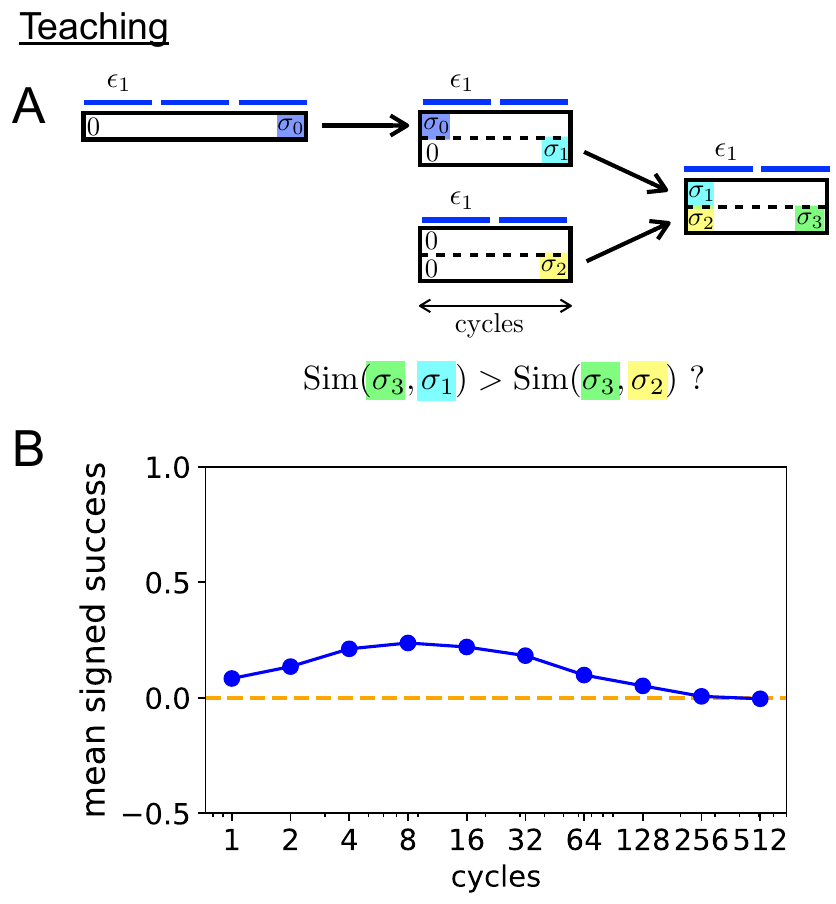}
\caption{Coupling to an adapted state. {\bf A.} A teacher is defined as a state that went through cycles of an environmental sequence. This state seeds a reactor that is coupled to another reactor. After a varying number of cycles in the same environmental sequence (x-axis in B), the final state of the second reactor defines an instructed student. Similarly, an untaught student is defined in the absence of a teacher. Finally, the instructed and uninstructed students compete in two coupled reactors. {\bf B.} Advantage in competition of being instructed as a function of the number of cycles in the second phase of the protocol. Interacting with a teacher is only beneficial if the training time is short enough so that a student without instruction does not entirely adapt on its own.
}\label{fig:teach}
\end{figure}

To further demonstrate that adaptation occurs beyond environmental response, we analyze the effect of interacting with a reactor containing a previously adapted state (Fig.~\ref{fig:teach}A). This teacher-student setting is implemented by a scheme where a ``teacher'' is first subjected to a number of environmental cycles, resulting in a state $\s_0$. This state seeds one of two coupled reactors, while the other is initially empty, corresponding to the initial condition $(\s_0,0)$. The final state in this initially empty reactor defines a student $\s_1$ that benefited from the teacher. As a control, a second student $\s_2$ is defined that does not benefit from a teacher, corresponding to the initial condition $(0,0)$. Finally, we seed two coupled reactors with $(\s_1, \s_2)$ to evaluate whether interacting with a teacher confers an adaptive advantage. The results confirm that this is indeed the case, provided teaching occurs over a sufficiently long but not too long time (Fig.~\ref{fig:teach}B). The advantage of teaching eventually decreases over time, because with enough time even a state that has not been instructed can adapt perfectly.

\subsection*{Coupling between unadapted systems}

\begin{figure}[t]
\centering
\includegraphics[width=.7\linewidth]{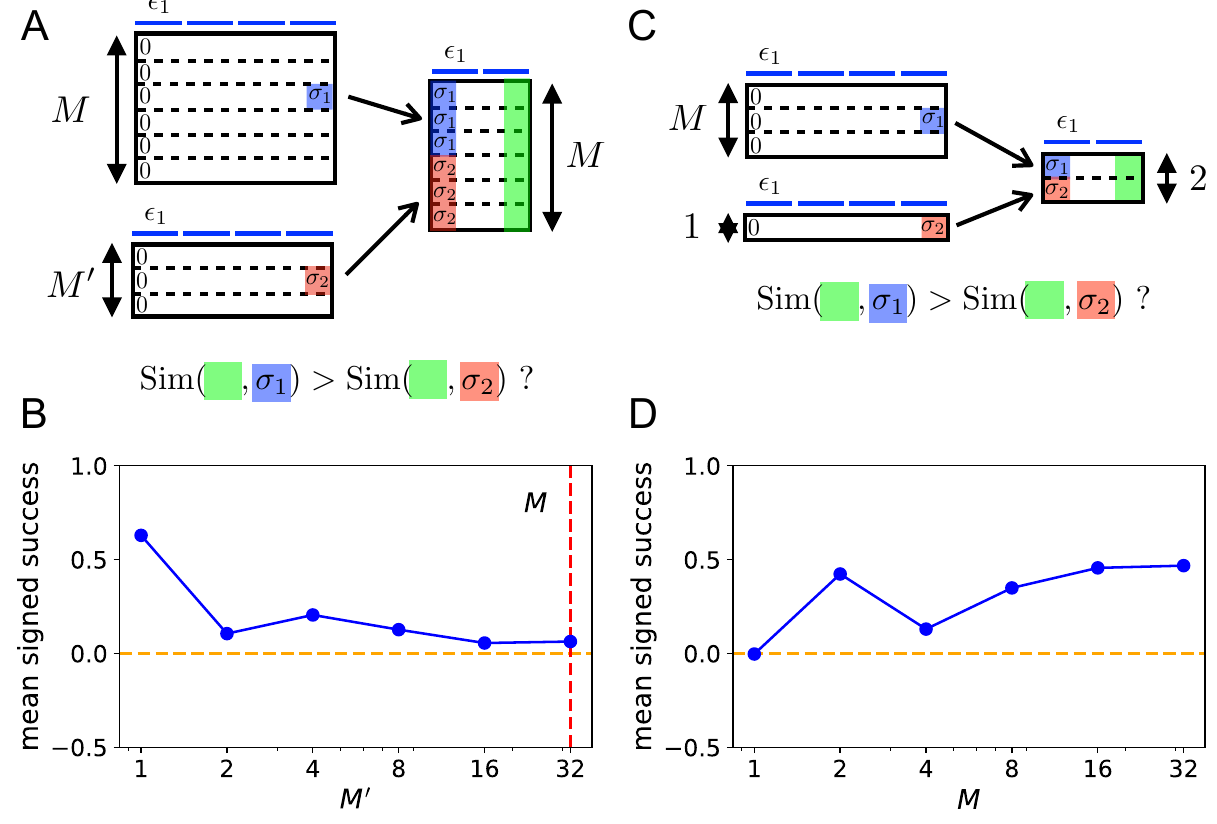}
\caption{Coupling between unadapted systems. {\bf A.} Competition in a system with $M$ reactors between a state $\s_1$ obtained in a system of $M$ reactors and a state $\s_2$ obtained in a system of $M'\leq M$ reactors. {\bf B.}~Results reporting that, on average, the state obtained in the context of more reactors wins ($M=32$ and varying $M'\leq M$). {\bf C.} Competition in two coupled reactors between a state obtained in the context of $M$ reactors and a state obtained in a single reactor. {\bf D.} Results reporting that, on average, the state obtained with more reactors wins (varying $M$).}\label{fig:M}
\end{figure}

We now generalize to $M>2$ coupled reactors arranged in a closed chain, as defined in \eqref{eq:multiple} and \eqref{eq:multconstraints}. We first assess whether interacting reactors provide an advantage by pitting two states, $\s_1$ and $\s_2$, originating from systems with different numbers of coupled reactors, $M$ and $M'\leq M$ respectively (Fig.~\ref{fig:M}A). Competition is set in a system of $M$ reactors by seeding $M/2$ consecutive reactors with $\s_1$ and the remainder with $\s_2$. Results demonstrate that larger systems yield better adapted states (Fig.~\ref{fig:M}B).

To further illustrate that (natural) selection under competition confers advantages beyond adaptation in a single reactor, we consider competition between a state obtained in multiple coupled reactors and one obtained in a single reactor, now with competition set in a system of $M=2$ reactors (Fig.~\ref{fig:M}C). The former generally outperforms the latter, further confirming that interactions between reactors promotes adaptation (Fig.~\ref{fig:M}D).

\section*{Conclusion}

Designing artificial systems that emulate the remarkable adaptive capabilities of living systems in simplified physical or chemical setups is a long-standing goal in science and engineering. The elaborate information processing seemingly required for complex adaptation has led researchers to draw inspiration from computers and modern artificial intelligence. This influence plays an important role in the field of physical learning, where neural network architectures are commonly taken as a starting point that is then simplified to permit physical implementation~\cite{stern2023learning}. However, alternative principles may prove more relevant for systems far simpler than the brain.

Here, we propose such an alternative perspective, inspired by biological adaptation in organisms without nervous systems. Even microorganisms lacking identifiable information-processing organs are indeed capable of adaptive behavior: after exposure to a new environment, they can outperform their peers not exposed to this environment when competing under the same conditions~\cite{Tagkopoulos.2008,Mitchell.2009,gershman2021reconsidering}. This observation leads to the speculation that the physicochemical processes  underlying reproduction might already be sufficient, by themselves, to enable an elementary form of learning. Our model not only demonstrates this possibility but shows that they may already give rise to remarkably elaborate forms of learning.

Our approach builds on a minimal model of coupled chemical reactions that can be realized using only elementary rigid blocks that associate and dissociate under thermal fluctuations~\cite{bunin2025evolutionary}. These reactions generate a large variety of out of equilibrium states which differ by the combination of molecular concentrations they maintain. This is achieved through a network of reactions with sparse, random and reciprocal inhibitions. As we showed previously, such dynamics reproduce core biological features: metabolism, heredity, and reproduction~\cite{bunin2025evolutionary}. Metabolism corresponds to maintaining specific out-of-equilibrium states driven by environmental inputs. Different influxes of reactants lead to distinct stable states, analogous to how a cell adopts different metabolic phenotypes depending on available nutrients. Heredity arises because stochastic perturbations, whether due to environmental fluctuations or molecular noise, can trigger transitions between states. Successive states being correlated, this creates a form of memory. Finally, reproduction refers to the ability of certain states to occupy increasingly larger parts of space over time, as they spread and expand at the expense of others.

Importantly, nothing in this setting is explicitly designed for learning, in particular by including the standard components of machine learning: there is no prescribed objective function, no a priori separation between ``neurons'' and ``weights'', and no external supervision. Yet, when exposed to complex environmental sequences, the system spontaneously develops adaptive states, whereby the new states harbor a reproductive advantage under the conditions in which they emerged, mirroring adaptation in simple living organisms.

How does this adaptation arise? A well-known mechanism of biological adaptation is natural selection. Our system indeed contains all ingredients necessary for Darwinian evolution by natural selection~\cite{bunin2025evolutionary}. However, we find that natural selection is not required for adaptation to occur as it can emerge in the absence of competition, in single well-mixed reactors where spatial heterogeneity, and thus reproduction, cannot take place. In this case, competition is only necessary to reveal adaptation: competition between states shows that the system has reached states with enhanced reproductive potential tailored to specific environments. This intrinsic form of learning resembles physiological adaptation, which occurs within organisms. When spatial heterogeneity is introduced, the system also exhibits collective learning. This form of learning resembles evolutionary adaptation, which occurs between individuals of a population. Finally, our system also demonstrates educated learning through interaction with another system that previously experienced the environment. 

The closest physical analogs to the present setting are disordered materials such as glasses, gels, granular materials, or mechanical metamaterials, whose hysteretic elements can encode and retain memory of past inputs~\cite{keim2019memory,stern2023learning,meulblok2025path}. In well-mixed conditions, our model maps formally to a spin glass on a random graph used to describe some of these systems (Methods). Such materials, however, lack reproduction and thus do not display adaptation in the sense of reproductive advantage. Conversely, existing proposals for replicating soft-matter systems usually lack the internal diversity required to produce complex, stimulus-specific responses~\cite{penrose1959self,von1986self,sakref2024design}. Notable exceptions include DNA-based assemblies and multi-type crystal systems that replicate and adapt through crystal growth and divisions~\cite{wang2011self,schulman2008crystals,schulman2012robust}. Yet, these designs rely on sophisticated engineering to reliably achieve replication. In contrast, our model exhibits adaptation without designing for it. 

More broadly, while previous studies suggest that designing either learning or replication is a difficult task by itself and that coupling them may be even more challenging, our model shows that by focusing on spatially extended systems capable of supporting a large diversity of out-of-equilibrium states, reproduction and learning can naturally arise and are intrinsically coupled. This result should help shed light on the remarkable adaptive abilities of simple organisms and suggest new approaches for designing artificial systems with lifelike properties.

\section*{Methods}

Here we describe how the present model with  discrete variables and discrete space is related to the model with continuous variables and space described previously ~\cite{bunin2025evolutionary}.

\subsection*{Continuous model}

The model consists of four types of molecular species, $A_i$, $B_i$, $C_i$, and $R_i$, with $i = 1, \dots, N$. The $A_i$ are the primary species of interest whose relative concentrations define the system's state: they are produced from the resources $R_i$ by means of the catalysts $C_i$; can inhibit the catalysts $C_j$ that mediate the formation of other $A_j$; and this inhibition can be prevented when $A_i$ binds to $B_i$. These reactions occur across three distinct timescales:
\begin{align}
& R_i + C_i \to A_i + C_i \quad && \text{(slow)}\label{eq:A}\\
& C_i + A_j \rightleftharpoons C_iA_j \quad && \text{(intermediate)}\label{eq:C}\\
& A_i + B_i \rightleftharpoons A_iB_i \quad && \text{(fast)} \label{eq:B}
\end{align}
The second reaction only occurs between certain pairs, $i\vdash\!\dashv j$, assumed to be reciprocally inhibitory, which are determined by the chemistry. Here we take, by design, that these pairs are sparse, with each $i$ participating in three such pairs chosen at random.

The reactions take place in an open reactor driven by resource influx, where each $R_i$ is injected at a given (possibly time-dependent) rate, and all species and complexes are diluted at a common rate $\delta$, both defined on the slow timescale. The rates are chosen such that species $C_i$ and $B_i$ are effectively chemostated. Although summarized here at the level of chemical reactions, we showed before how these reactions can arise from simple building blocks that undergo Browinian motion and thermal association and dissociation~\cite{bunin2025evolutionary}.

\subsection*{Coarse-graining}

We first consider a single well-mixed reactor of large volume. The separation of timescales allows for coarse-graining into effective reactions that simplify the mathematical description. Starting on the fastest timescale with \eqref{eq:B}  and assuming strong interaction between $A_i$ and $B_i$, the concentration of free $A_i$ is the excess not bound to $B_i$, i.e. $[A_i]_{\rm free}=[A_i]_{\rm tot}-[B_i]_{\rm tot}$ if $[A_i]_{\rm tot}>[B_i]_{\rm tot}$, and $[A_i]_{\rm free}=0$ otherwise, also written
\beq
[A_i]_{\rm free}=\max(0, [A_i]_{\rm tot}-[B_i]_{\rm tot})
\eeq
Next, assuming that \eqref{eq:C} equilibrates on an intermediate timescale, $[C_iA_j]=\Gamma_{ij} [C_i]_{\rm free}[A_j]_{\rm free}$, where $\Gamma_{ij}$ is the association constant. As a consequence, the concentration of free $C_i$, which satisfies the conservation law
\beq
[C_i]_{\rm tot}=[C_i]_{\rm free}+\sum_{j\vdash\!\dashv i}[C_iA_j]
\eeq
is given by
\beq
[C_i]_{\rm free}=\frac{[C_i]_{\rm tot}}{1+\sum_{j\vdash\!\dashv i}\Gamma_{ij}[A_j]_{\rm free}}
\eeq
Finally, on the slowest timescale, combining \eqref{eq:A} with the injection of $R_i$ and the dilution of all complexes involving $A_i$, the total concentration of $A_i$ follows
\beq
\partial_t[A_i]_{\rm tot}=\alpha_i [C_i]_{\rm free}[R_i]-\delta [A_i]_{\rm tot}
\eeq
All together, assuming all types to be equivalent with $\L_i=\a_i[R_i][C_i]_{\rm tot}$, $\Gamma_{ij}=\G$ and $[B_i]_{\rm tot}=B$, the dynamics of the $n_i= [A_i]_{\rm tot}$ is of the form
\beq\label{eq:basic}
\partial_t n_i(t)=F_i(n(t),\L)
\eeq
with
\beq
F_i(n,\L)=\L_i c_i(n)-\delta n_i
\eeq
and 
\beq
c_i(n)=\frac{1}{1+\G\sum_{j\vdash\!\dashv i}\max(0,n_j-B)}.
\eeq

\subsection*{Out-of-equilibrium steady states}

A simple example chemistry is $N=2$ with reciprocal inhibition, $1\vdash\!\dashv 2$. This system admits two steady states, one obtained starting from a high concentration of $A_1$ which inhibits $C_2$ and keeps $A_2$ at low concentration; the other has the roles of $A_1$ and $A_2$ reversed. The role of $B_i$ is to stabilize the inactive state of $A_i$.

For general $N$, up to $2^N$ combinations are possible, where each $A_i$ is either inhibited or actively produced. However, not all are realizable steady states: if $A_i$ and $A_j$ inhibit each other, at most one can be active at a given time. 
Thus, steady states are characterized by a set of active nodes such that no two active nodes are connected, and no inactive node is connected exclusively to inactive nodes. These states correspond to solutions of a well-known graph-theoretic problem called independent sets, here applied to the inhibitory graph where each $i$ is a node and where edges are defined by $i\vdash\!\dashv j$. When each $i$ is inhibited by three other $j$ chosen at random, the number of steady states grows exponentially with $N$.

\subsection*{Diffusion}

Space plays an important role in the model, as it enables growth. In a continuous setting, it is described by extending \eqref{eq:basic} to include diffusion,
\beq\label{eq:diff}
\p_tn_i(x,t)=F_i(n(x,t),\L)+D\nabla^2 n_i(x,t)
\eeq

\subsection*{Noise}

Noise also plays an important role, as it enables random transitions between steady states.
It is induced by fluctuations in the number of molecules. We describe it with chemical Langevin equations of the form of \eqref{eq:rde} with
\beq
{\rm noise}(x,t)=T \sqrt{c_i(n(x,t))+\d n_i(x,t)}\ \xi_i(x,t)+T\sqrt{2D}\partial_x\left(\sqrt{n_i(x,t)}\zeta_i(x,t)\right)
\eeq
where $\xi_i(x,t)$ and $\zeta_i(x,t)$ are Gaussian white noises, and where $T$ is a ``temperature'' representing the strength of the noise.

\subsection*{Discrete variables}

At low enough noise, the dynamics in a single reactor switches between compositions resembling deterministic steady states (see Fig.~\ref{fig:scheme}A). These states are described using discrete variables $\sigma_i$, where $\sigma_i=1$ indicates that $A_i$ is active and $\sigma_i=0$ that it is inhibited.

We showed previously~\cite{bunin2025evolutionary} that the out-of-equilibrium transitions between these discrete steady states satisfies detailed-balance, namely it is equivalent to \emph{equilibrium dynamics} described by Metropolis rates, with an energy function
\beq
E[\sigma]=-\sum_{i=1}^N(\Lambda_i/\delta)\sigma_i
\eeq
and the effective temperature $T=1/\sqrt{\Omega}$, where $\Omega$ is the reactor volume. This energy is defined under the constraints that $\sigma_i=0$ whenever $\Lambda_i=0$, and $\sigma_i\sigma_j=0$ whenever $i\vdash\!\dashv j$ (the definition of realizable steady states). This mapping to an equilibrium process, which is not possible for arbitrary out-of-equilibrium dynamics, relies crucially on the reciprocity of inhibitory interactions.

\subsection*{Discrete space}

We also simplify the model by dividing space into distinct coupled reactors, where the concentrations in each reactor are uniform (well-mixed), and between the reactors elements diffuse at a rate $\phi$. If this rate is small and if $\S(x)$ denotes the set of reactors coupled to reactor $x$, the reaction-diffusion equation, \eqref{eq:diff}, becomes
\beq
\partial_t n_{i}(x,t)= \L_i(t)c_i(n(x,t))-\delta n_{i}(x,t)+\phi\sum_{y\in\S(x)}(n_i(y,t)-n_i(x,t))\;.
\eeq
With discrete variables $\s_i(x)$ describing the state in each reactor $x$, and for small $\phi$, transitions in each reactor effectively occur with a modified influx $\tilde\L_{i}(x)$ that includes the diffusion from neighboring reactors $\S(x)$, 
\beq
\tilde\L_{i}(x)=\left(1+\phi\sum_{y\in\S_x}(\s_{i}(y)-\s_{i}(x))\right)\L_i
\eeq
When all reactors are coupled to the same number $r$ of other reactors, the joint dynamics of states is governed by the energy
\beq
E[\s]=-(1-r\phi)/\d\sum_{i,x}\s_{i}(x)-\phi/\d\sum_{i,x,y\in\S(x)}\s_{i}(x)\s_{i}(y)
\eeq
This is the form that we use in \eqref{eq:multiple}, up to a multiplicative factor. The value used for $J$ in our simulations, $J=0.5$, is not so small, but results are qualitatively similar to the continuous model, on which Fig.~\ref{fig:illustr}B is based.

\subsection*{Spin-glass interpretation}

The model in a single reactor, as defined by \eqref{eq:Constraints} and \eqref{eq:Energy}, corresponds to a spin-glass model on a sparse graph, consisting of particles with hard-core repulsion. This model, known as the hard-core model, is also studied as a combinatorial optimization problem called vertex cover. On random graphs with low connectivity, it is known to exhibit a continuous (full-replica symmetry-breaking) glass transition at low temperature~\cite{barbier2013hard}. From this point of view, \eqref{eq:J} describes real replicas of hard-core models coupled along a one-dimensional line. To our knowledge, this spatial extension has not been previously studied in the spin-glass literature.

\end{document}